# Transverse spin and surface waves in acoustic metamaterials

Konstantin Y. Bliokh[1,2] and Franco Nori[1,3]

[1]*Theoretical Quantum Physics Laboratory, RIKEN Cluster for Pioneering Research, Wako-shi, Saitama 351-0198, Japan*
[2]*Nonlinear Physics Centre, RSPE, The Australian National University, ACT 0200, Australia*
[3]*Physics Department, University of Michigan, Ann Arbor, Michigan 48109-1040, USA*

We consider spin angular momentum density in inhomogeneous acoustic fields: evanescent waves and surface waves at interfaces with negative-density metamaterials. Despite being purely longitudinal (curl-free), acoustic waves possess intrinsic *vector* properties described by the velocity field. Motivated by the recent description and observation of the spin properties in elastic and acoustic waves, we compare these properties with their well-known electromagnetic counterparts. Surprisingly, both the transverse spin of evanescent waves and the parameters of surface waves are very similar in electromagnetism and acoustics. We also briefly analyze the important role of dispersion in the description of the energy and spin densities in acoustic metamaterials.

## 1. Introduction

It is well known that circularly-polarized electromagnetic waves carry intrinsic angular momentum (AM) corresponding to the spin-1 of photons in the quantum description [1–4]. For plane waves or paraxial beams, this spin AM is longitudinal, i.e., directed along the wavevector. In the past several years, it was shown that an unusual *transverse spin AM* density can appear in *inhomogeneous* optical fields, such as evanescent waves or non-paraxial interference fields [5–10] (for reviews, see [4,11]). This local transverse spin can arise even for transverse-electric (TE) or transverse-magnetic (TM) waves, usually considered as linearly-polarized, due to the presence of a non-zero longitudinal field component, which generates an effectively-elliptical polarization in the propagation plane. The transverse optical spin has found interesting applications in highly efficient spin-direction coupling and routing [12–17], important for spin-based optical and quantum networks (see [18] for a review).

Recently, it was shown that nontrivial spin AM density also naturally appears for waves in elastic media [19]. Importantly, both transversal (i.e., divergence-free, similar to electromagnetic) and longitudinal (i.e., curl-free, acoustic) modes equally contribute to the local spin AM density and can generate transverse spin in inhomogeneous elastic waves. This is in contrast to the popular belief that longitudinal phonons are spin-0 particles, which cannot have any nontrivial vector properties. Moreover, an experimental observation of the transverse spin AM density in purely longitudinal inhomogeneous acoustic waves was reported very recently [20].

In this work, motivated by the studies [19,20], we analyze the equations for longitudinal acoustic waves and show that the spin AM density and transverse spin naturally appear there from the circular motion of the medium particles. We argue that despite their longitudinal character, the acoustic waves *cannot* be considered as *scalar* waves. These are still *vector* waves, described by the vector velocity and scalar pressure fields. Moreover, we show that the acoustic wave equations exhibit close similarity with the Maxwell equations in isotropic optical media. As examples, we consider evanescent acoustic waves, carrying transverse spin AM, and surface acoustic modes [21,22] at interfaces with acoustic metamaterials (negative-parameter media) [23–25]. Most of their features demonstrate an exact analogy with TM surface modes [26,27] (such as surface plasmon-polaritons [28]) in Maxwell electromagnetism. Finally, using an analogy with electromagnetic waves in dispersive media [29–31], we analyze the important role of the dispersion in determining the energy and spin densities in acoustic metamaterials.

## 2. Transverse spin in evanescent acoustic waves

We start with the equations for monochromatic acoustic waves of frequency $\omega$ in a dense medium:

$$\nabla \cdot \mathbf{v} = i\beta\omega P, \quad \nabla P = i\rho\omega \mathbf{v}. \qquad (1)$$

Here the variables are: the complex velocity $\mathbf{v}(\mathbf{r})$ and the pressure $P(\mathbf{r})$ fields, while the real-valued medium parameters are: the mass density $\rho$ and the compressibility $\beta = 1/B$ ($B$ is the bulk modulus).

Equations (1) support only longitudinal (i.e., curl-free) waves, which follows from the second Eq. (1): $\nabla \times \mathbf{v} = 0$. Importantly, these are *not* scalar waves. Indeed, the velocity $\mathbf{v}$ determines *vector* properties of acoustic waves, even though these are longitudinal. For plane waves with the wavevector $\mathbf{k}$, $\nabla \to i\mathbf{k}$, the dispersion relation and the "longitudinality" condition follow from Eqs. (1):

$$\omega^2 = k^2 c^2 \equiv \frac{k^2}{\rho\beta}, \quad \mathbf{k} \times \mathbf{v} = 0. \qquad (2)$$

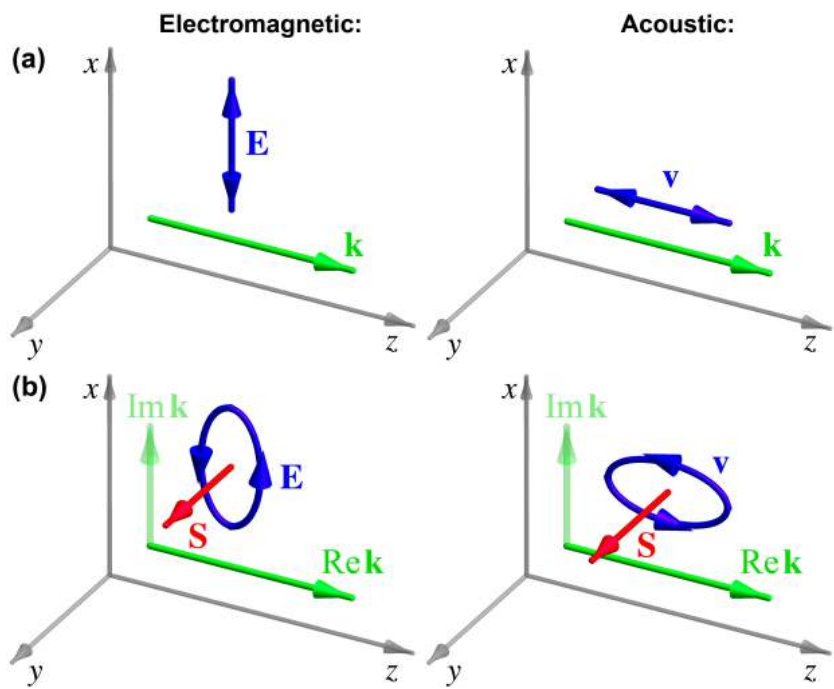

**Fig. 1.** Appearance of the transverse spin AM in transversal (electromagnetic) and longitudinal (acoustic) waves. **(a)** Plane waves of these types have linear polarizations orthogonal and parallel to the wavevector, respectively. **(b)** Evanescent waves with orthogonal real and imaginary parts of the wavevector inevitably have elliptical in-plane polarization because of the transversality ($\mathbf{k} \cdot \mathbf{E} = 0$) and longitudinality ($\mathbf{k} \times \mathbf{v} = 0$) conditions. These elliptical polarizations generate the transverse spin $\mathbf{S}$.

Evanescent waves can be presented as plane waves with a *complex* wavevector $\mathbf{k} = \operatorname{Re}\mathbf{k} + i\operatorname{Im}\mathbf{k}$ [4,6,16]. Considering an evanescent wave propagating along the $z$-axis and decaying in the $x$-direction, $\mathbf{k} = k_z \bar{\mathbf{z}} + i\kappa \bar{\mathbf{x}}$ (the overbar indicates unit vectors), the dispersion relation (2) becomes $\omega^2 = (k_z^2 - \kappa^2)c^2$ (note that $k^2 = \mathbf{k}\cdot\mathbf{k} \neq |\mathbf{k}|^2$), while the longitudinality condition yields

$$v_y = 0, \qquad k_z v_x - i\kappa v_z = 0. \tag{3}$$

Thus, the velocity field has two components with the $\pi/2$ phase difference, exactly as in electromagnetic evanescent waves [4,6,16,18]. The only difference is that the "main" component of the transversal (i.e., divergence-free, $\nabla \cdot \mathbf{E} = i\mathbf{k}\cdot\mathbf{E} = 0$) electric field of a TM-polarized evanescent electromagnetic wave is $E_x$ and $E_z = -i(\kappa/k_z)E_x$, while for the longitudinal acoustic waves the main velocity component is $v_z$ and $v_x = i(\kappa/k_z)v_z$, as shown in Fig. 1.

As it is known for electromagnetic waves [4–11,18], and was very recently shown for elastic and acoustic waves [19,20], the orthogonal imaginary component in the vector field produces *rotation* of this vector field, i.e., elliptical polarization and *spin AM* in the wave, Fig. 1. Entirely similar to the TE or TM electromagnetic waves, this spin is *transverse* in evanescent acoustic waves, i.e., directed along the $y$-axis orthogonal to the propagation $(x,z)$-plane determined by the $\mathbf{k}$-vector.

The general expression for the time-averaged spin AM density $\mathbf{S}$ can be written by noticing that the medium particles with mass density $\rho$ experience complex displacements $\mathbf{a}$, $\mathbf{v} = -i\omega\mathbf{a}$:

$$\mathbf{S} = \frac{\rho}{2}\operatorname{Re}(\mathbf{a}^* \times \mathbf{v}) = \frac{\rho}{2\omega}\operatorname{Im}(\mathbf{v}^* \times \mathbf{v}). \tag{4}$$

Notably, the entirely similar contribution of oscillating electrons in optical media provides the material contribution to the electromagnetic spin AM [31]. For the normalization of the spin (4), we introduce the time-averaged energy density of acoustic waves, which follows from the acoustic analogue of the Poynting theorem for Eqs. (1):

$$W = \frac{1}{4}\left(\rho|\mathbf{v}|^2 + \beta|P|^2\right). \tag{5}$$

Substituting Eqs. (1), (3), and the dispersion relation for the evanescent wave considered above into Eqs. (4) and (5), we find that the normalized spin AM density in the acoustic evanescent wave is

$$\frac{\omega\mathbf{S}}{W} = \frac{2\kappa}{k_z}\bar{\mathbf{y}} = \frac{2\operatorname{Re}\mathbf{k} \times \operatorname{Im}\mathbf{k}}{(\operatorname{Re}\mathbf{k})^2}. \tag{6}$$

This expression is entirely analogous to the transverse spin density in electromagnetic evanescent waves in vacuum [4,6,18] *up to a factor of* 2. This difference originates from the fact that in electromagnetism both the spin AM and energy are equally distributed between the electric and magnetic ($\mathbf{E}$ and $\mathbf{H}$) vector degrees of freedom [4,6,30,31]. In contrast, acoustic waves are described by one vector ($\mathbf{v}$) and one scalar ($P$) fields. Therefore, while energy is equally distributed between these fields, Eq. (5), the spin AM is fully concentrated in the vector velocity field, Eq. (4). This leads to the above difference in the relative coefficients for electromagnetic and acoustic quantities. Since $\kappa < k_z$, the absolute value of the transverse spin density "per phonon" (assuming the energy density $W = \hbar\omega$ per phonon) is restricted by the value of $2\hbar$.

It is important to emphasize that although the acoustic spin AM (4) originates from the mechanical circular motion of microscopic particles of the acoustic medium, it is a purely *intrinsic* and local property of the *macroscopic* wave field. Indeed, the rotation of the vector velocity field **v** occurs at a given point **r** and does not involve its neighborhood. This is entirely similar to the local rotation of the electric field **E** in an elliptically polarized electromagnetic wave and is in contrast to the previously-considered acoustic *orbital* AM related to acoustic *vortices* [32–34]. In the vortex case, the orbital AM originates from the azimuthal phase gradients of the *scalar* wave field in the vicinity of the vortex core.

## 3. Surface waves at interfaces with acoustic metamaterials

We now consider an important example of acoustic modes involving evanescent waves. These are *surface waves* that appear at interfaces between two media [21,22]. Akin to electromagnetic surface waves (e.g., surface plasmon-polaritons), which exist at interfaces between media with positive and negative parameters [26–28], acoustic surface modes appear at interfaces between media with positive and negative densities $\rho$ [21,22]. However, while natural optical media can have negative parameters (e.g., negative permittivity in metals), negative acoustic parameters are present only in *metamaterials*, which consist of many microscopic (subwavelength) acoustic resonators [23–25]. The presence of these resonators yields strong *dispersion* of the effective medium parameters, $\rho(\omega)$ and $\beta(\omega)$, and can produce negative values at certain frequencies. In this section, we do not consider these dispersion and metamaterial aspects, and only assume that the medium parameters can take on any real values (losses are neglected) at the given frequency $\omega$.

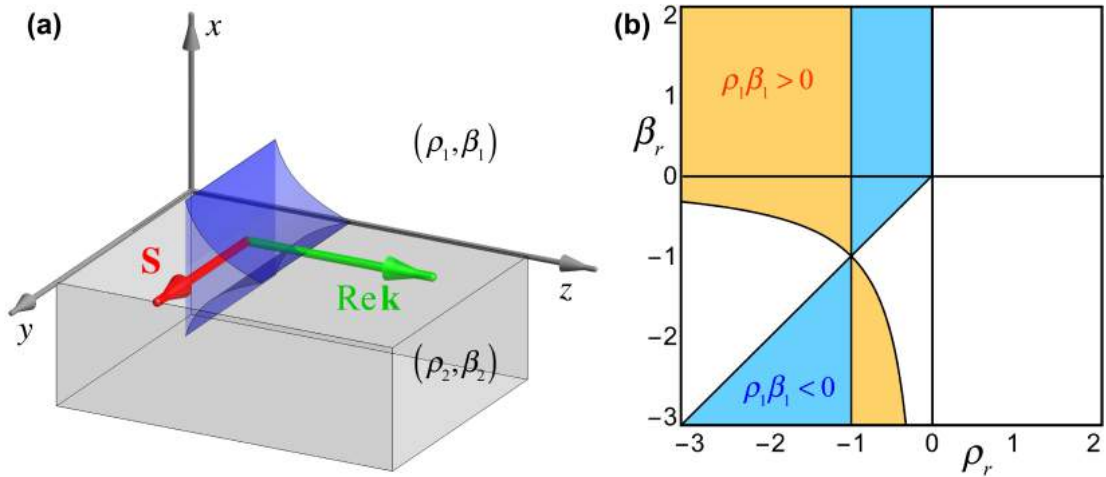

**Fig. 2. (a)** Schematics of a surface wave at the interface between two acoustic media [21,22]. **(b)** Phase diagram showing the zones of existence of surface modes, Eqs. (7) and (8), with real frequency $\omega$ and propagation constant $k_z$ in the parameter plane $(\rho_r, \beta_r)$. Two cases $\rho_1 \beta_1 > 0$ (transparent medium 1) and $\rho_1 \beta_1 < 0$ (non-transparent medium 1) are shown in blue and yellow, respectively. This phase diagram exactly coincides with the diagram for TM-polarized surface electromagnetic waves depending on the $(\varepsilon, \mu)$ parameters [26,27].

The geometry of the problem in shown in Fig. 2(a). The surface mode at the $x = 0$ interface consist of two evanescent waves with the same propagation constant $k_z$ and opposite-sign decay constants $\kappa_1 > 0$ and $\kappa_2 < 0$. Solving Eqs. (1) supplied with the boundary conditions (the continuity of $P$ and $v_x$) yields the condition of the surface-wave existence [21]:

$$\frac{\kappa_1}{\rho_1} = \frac{\kappa_2}{\rho_2}. \tag{7}$$

Thus, surface acoustic waves can exist only at interfaces between positive- and negative-density media. Supplying Eq. (7) with the dispersion relations for the evanescent waves in the two media, $\omega^2 = c_{1,2}^2 \left( k_z^2 - \kappa_{1,2}^2 \right)$, $c_{1,2}^2 = 1/\left( \rho_{1,2} \beta_{1,2} \right)$, we find the dispersion characteristics of the surface mode:

$$k_z^2 = \kappa_1^2 \frac{\rho_r \left( \beta_r - \rho_r \right)}{\rho_r \beta_r - 1}, \quad \rho_1 \beta_1 \omega^2 = \kappa_1^2 \frac{\left( 1 - \rho_r^2 \right)}{\rho_r \beta_r - 1}, \tag{8}$$

where $\rho_r = \rho_2 / \rho_1$ and $\beta_r = \beta_2 / \beta_1$ are the relative parameters of the two media.

Remarkably, Eqs. (7) and (8) are *precisely equivalent* to the equations for the TM-polarized electromagnetic surface waves (including surface plasmon-polaritons at metal-dielectric interfaces) with the substitution $(\rho, \beta) \leftrightarrow (\varepsilon, \mu)$ ($\varepsilon$ and $\mu$ being the permittivity and permeability of the optical media) [26,27]. The zones of the existence of surface modes with real $\omega$ and $k_z$, given by Eqs. (8), in the $(\rho_r, \beta_r)$ parameter plane are depicted in Fig. 2(b). This phase diagram exactly coincides with its TM-electromagnetic counterpart, showing two mutually-complementary zones for the cases $\rho_1 \beta_1 > 0$ (transparent first medium) and $\rho_1 \beta_1 < 0$ (non-transparent first medium) [26,27].

We are now at a position to calculate the transverse spin (4) and (6) for the acoustic surface wave. The spin and energy densities (4) and (5) in the evanescent waves in the two media are given by

$$S_y = \rho_{1,2} \frac{\kappa_{1,2}}{\omega k_z} \left| v_{z1,2} \right|^2 = \frac{k_z \kappa_{1,2}}{\omega^3 \rho_{1,2}} \left| P_{1,2} \right|^2, \quad W = \frac{\rho_{1,2}}{2} \left| v_{z1,2} \right|^2 = \frac{k_z^2}{2 \omega^2 \rho_{1,2}} \left| P_{1,2} \right|^2, \tag{9}$$

where we used $k_z P = \rho \omega v_z$ from the second Eq. (1). Since the pressure is continuous across the interface, one can write $\left| P_{1,2} \right|^2 = A^2 \exp\left( -2 \kappa_{1,2} x \right)$, where $A$ is a constant amplitude. From here and Eqs. (9), we calculate the ratio of integral values of the spin and energy in the surface wave:

$$\frac{\omega \langle S_y \rangle}{\langle W \rangle} = \frac{2 \kappa_1 \kappa_2 \left( \rho_2 - \rho_1 \right)}{k_z \left( \kappa_2 \rho_2 - \kappa_1 \rho_1 \right)} = -\frac{2}{1 + \rho_r} \sqrt{\frac{\rho_r \left( \rho_r \beta_r - 1 \right)}{\beta_r - \rho_r}}, \tag{10}$$

where $\langle ... \rangle \equiv \int_{-\infty}^{\infty} ... dx$, and we used Eqs. (7) and (8) to obtain the final expression (10).

Note that since both $\kappa$ and $\rho$ have opposite signs in the two media, the spin density $S_y$ has the *same* sign in the two media, while the energy density $W$ has *opposite* signs, Eqs. (9). As a result, the integral energy $\langle W \rangle$ and the normalized integral spin (10) can take on arbitrary real values. The negative energy density appears here because we did not take into account the dispersion (similar to the negative electric energy density $\varepsilon|\mathbf{E}|^2$ in a hypothetic nondispersive optical medium with $\varepsilon < 0$).

## 4. Modifications by dispersion

As we mentioned, the negative density $\rho$ can appear only in highly dispersive acoustic metamaterials with frequency-dependent parameters $\rho(\omega)$ and $\beta(\omega)$. Furthermore, as it is known from electrodynamics, the general expressions for the energy and spin densities considerably change their forms in the presence of dispersion. Namely, as it was shown by Brillouin for the energy density [29] and recently for the spin AM density and other dynamical characteristics [30,31,35], the dispersion requires substituting the coefficients $(\varepsilon,\mu) \to (\tilde{\varepsilon},\tilde{\mu}) \equiv (\varepsilon,\mu) + \omega\, d(\varepsilon,\mu)/d\omega$ in all bi-linear forms for electromagnetic waves. Importantly, the modified quantities $(\tilde{\varepsilon},\tilde{\mu})$ are always *positive*, and this guarantees that the bi-linear energy form is positive definite.

The acoustic energy density in a dispersive medium can be derived similarly to the Brillouin electromagnetic energy density [29]. Namely, we introduce auxiliary fields $\mathbf{M}$ (momentum density) and $N$, which are related to the velocity and pressure fields in the frequency domain as

$$\mathbf{M} = \rho\mathbf{v}, \quad N = \beta P. \tag{11}$$

In the time domain, these fields determine the acoustic analogue of the Poynting theorem [following from the time-domain versions of Eqs. (1)]:

$$\mathbf{v} \cdot \frac{\partial \mathbf{M}}{\partial t} + P\frac{\partial N}{\partial t} + \nabla \cdot (P\mathbf{v}) = 0. \tag{12}$$

Repeating the electromagnetic textbook considerations [29] for these quantities, we find the time-averaged acoustic energy density for monochromatic fields in a dispersive medium:

$$\tilde{W} = \frac{1}{4}\left(\tilde{\rho}|\mathbf{v}|^2 + \tilde{\beta}|P|^2\right), \quad (\tilde{\rho},\tilde{\beta}) = (\rho,\beta) + \omega\frac{d(\rho,\beta)}{d\omega}. \tag{13}$$

Because of the very close analogy between the electromagnetic and acoustic problems, it is natural to assume that the acoustic spin AM density is also modified similarly to the electromagnetic one in the presence of dispersion [30,31,35]:

$$\tilde{\mathbf{S}} = \frac{\tilde{\rho}}{2\omega}\operatorname{Im}(\mathbf{v}^* \times \mathbf{v}). \tag{14}$$

The dispersive corrections can considerably modify the quantities under consideration. For example, consider the resonant Lorentz-type dispersion of the density [21]: $\rho(\omega) = \rho_0 \omega_0^2 / (\omega_0^2 - \omega^2)$, where $\rho_0 > 0$ is a constant, and $\omega_0$ is the resonant frequency of microresonators in the metamaterial. The density becomes negative above the resonant frequency: $\rho(\omega) < 0$ for $\omega > \omega_0$. However, the modified density parameter is always positive: $\tilde{\rho} = \rho_0 \omega_0^2 (\omega_0^2 + \omega^2)/(\omega_0^2 - \omega^2)^2 > 0$, which ensures that the energy density (13) is positive definite.

The local ratio of the spin and energy densities in an evanescent wave, Eq. (6), is also modified by the presence of the dispersion. Using the definitions (13) and (14), we find:

$$\frac{\omega \tilde{S}_y}{\tilde{W}} = \frac{4\tilde{\rho}\beta\kappa k_z}{\tilde{\rho}\beta(k_z^2+\kappa^2)+\rho\tilde{\beta}(k_z^2-\kappa^2)}. \tag{15}$$

The normalized integral value of the spin AM of a surface wave, Eq. (10), can also be calculated straightforwardly with the dispersion-modified definitions (13) and (14). We do not provide here the cumbersome resulting equations and only note that the dispersive analogue of the energy density (9) now becomes positive in both media, while the spin density has opposite signs in the two media (because $\tilde{\rho}$ is always positive, while $\kappa$ has opposite signs in the two media).

## 5. Conclusions

We have considered the spin AM density in longitudinal (curl-free) acoustic waves. The spin AM is produced by the local elliptical motion of the particles of the medium, which is described by the velocity field $\mathbf{v}$. Akin to the transversality condition ($\nabla \cdot \mathbf{E} = 0$) for electromagnetic waves, the "longitudinality" condition for acoustic waves ($\nabla \times \mathbf{v} = 0$) generates the elliptical motion and transverse spin AM in inhomogeneous acoustic fields. As it is known for electromagnetic waves [4–11,18,36] and was very recently shown for elastic and acoustic waves [19,20], this phenomenon is generic for any inhomogeneous fields, including interference fields, tightly focused beams, etc.

|  | **Electromagnetism** | **Acoustics** |
|---|---|---|
| Fields | electric $\mathbf{E}$, magnetic $\mathbf{H}$ | velocity $\mathbf{v}$, pressure $P$ |
| Constraints | $\nabla \cdot \mathbf{E} = 0$, $\nabla \cdot \mathbf{H} = 0$ | $\nabla \times \mathbf{v} = 0$ |
| Medium parameters | permittivity $\varepsilon$, permeability $\mu$ | density $\rho$, compressibility $\beta$ |
| Energy density | $\frac{1}{4}\left(\varepsilon|\mathbf{E}|^2 + \mu|\mathbf{H}|^2\right)$ | $\frac{1}{4}\left(\rho|\mathbf{v}|^2 + \beta|P|^2\right)$ |
| Spin AM density | $\frac{1}{4}\left[\varepsilon \operatorname{Im}(\mathbf{E}^* \times \mathbf{E}) + \mu \operatorname{Im}(\mathbf{H}^* \times \mathbf{H})\right]$ | $\frac{1}{2}\rho \operatorname{Im}(\mathbf{v}^* \times \mathbf{v})$ |
| Transverse spin density in an evanescent wave | $\dfrac{\omega \mathbf{S}}{W} = \dfrac{\operatorname{Re}\mathbf{k} \times \operatorname{Im}\mathbf{k}}{(\operatorname{Re}\mathbf{k})^2}$ | $\dfrac{\omega \mathbf{S}}{W} = 2\dfrac{\operatorname{Re}\mathbf{k} \times \operatorname{Im}\mathbf{k}}{(\operatorname{Re}\mathbf{k})^2}$ |
| Surface waves | $\dfrac{\varepsilon_2}{\varepsilon_1} < 0$ (TM), $\dfrac{\mu_2}{\mu_1} < 0$ (TE) | $\dfrac{\rho_2}{\rho_1} < 0$ (TM-like) |
| Dispersive corrections | $(\varepsilon, \mu) \to (\tilde{\varepsilon}, \tilde{\mu})$ | $(\rho, \beta) \to (\tilde{\rho}, \tilde{\beta})$ |

**Table I.** Comparison of electromagnetic and acoustic quantities and properties.

We have considered evanescent acoustic waves and surface waves at interfaces between positive- and negative-density materials. In both cases, the acoustic equations exhibit very close similarity with Maxwell equations. Surprisingly, the transversal spin-1 and longitudinal spin-0 nature of photons and phonons, respectively, does not lead to any crucial difference between these two problems. The main difference between acoustic and Maxwell problems is that the electromagnetic waves are described by two vector degrees of freedom (electric and magnetic fields), which both equally contribute to all electromagnetic properties, while acoustic waves are characterized by one vector (velocity) and one scalar (pressure) fields, with only one of these producing spin AM. The comparison of electromagnetic and acoustic quantities and properties considered in this work is summarized in Table I.

Two important conclusions can be drawn from this comparison. First, the transverse-spin properties of inhomogeneous vector fields are independent from the usual spin of the corresponding quantum particles (e.g., spin-1 of photons and spin-0 of phonons). In fact, the usual-spin properties of photons correspond to *plane waves* and the polarization in the plane orthogonal to the wavevector (helicity), while the transverse spin is independent of this polarization and appears even for "linearly-polarized" TE and TM waves due to the longitudinal field component, Fig. 1. Second, acoustic waves have intrinsic *vector* properties, which cannot be described within the scalar wave equation. Indeed, it is known that the second-order scalar wave equation for the pressure field $P$ follows from the set (1) of two first-order equations. However, the scalar wave equation contains only a single combined parameter $(\rho\beta)$ (characterizing the sound-wave velocity), while, for example, the parameters and phase diagrams of surface acoustic waves depend on two parameters $\rho$ and $\beta$ separately, as seen from Fig. 2(b) and Eqs. (7) and (8). Therefore, the system of Eqs. (1), involving the vector velocity field $\mathbf{v}$, is essentially required for such problems.

Finally, similar to Maxwell electromagnetism, surface waves and negative-parameter media (acoustic metamaterials) are always accompanied by considerable *dispersion*. This can significantly modify definitions of the energy and spin densities in such media. Employing the analogy with electrodynamics, we have put forward the dispersion-modified forms of the energy and spin densities in acoustic metamaterials. A more rigorous derivation of these quantities requires a microscopic theory taking into account the internal degrees of freedom of microresonators, which constitute the metamaterial. This is an important problem for future study.

**Acknowledgements:** We are grateful to A.Y. Bekshaev, Y.P. Bliokh, M. Lein and D. Leykam for fruitful discussions. This work was partially supported by MURI Center for Dynamic Magneto-Optics via the Air Force Office of Scientific Research (AFOSR) (FA9550-14-1-0040), Army Research Office (ARO) (Grant No. Grant No. W911NF-18-1-0358), Asian Office of Aerospace Research and Development (AOARD) (Grant No. FA2386-18-1-4045), Japan Science and Technology Agency (JST) (Q-LEAP program, ImPACT program, and CREST Grant No. JPMJCR1676), Japan Society for the Promotion of Science (JSPS) (JSPS-RFBR Grant No. 17-52-50023, and JSPS-FWO Grant No. VS.059.18N), RIKEN-AIST Challenge Research Fund, the John Templeton Foundation, and the Australian Research Council.

# References

1. J.H. Poynting, "The wave-motion of a revolving shaft, and a suggestion as to the angular momentum in a beam of circularly-polarized light", *Proc. R. Soc. Lond. Ser. A* **82**, 560 (1909).
2. R.A. Beth, "Mechanical detection and measurement of the angular momentum of light", *Phys. Rev.* **50**, 115 (1936).
3. L. Allen, S.M. Barnett, and M.J. Padgett, *Optical Angular Momentum* (IoP Publishing, Bristol, 2003).
4. K.Y. Bliokh and F. Nori, "Transverse and longitudinal angular momenta of light," *Phys. Rep.* **592**, 1 (2015).
5. K.Y. Bliokh, F. Nori, "Transverse spin of a surface polariton", *Phys. Rev. A* **85**, 061801(R) (2012).
6. K.Y. Bliokh, A.Y. Bekshaev, and F. Nori, "Extraordinary momentum and spin in evanescent waves", *Nature Commun.* **5**, 3300 (2014).
7. A. Canaguier-Durand and C. Genet, "Transverse spinning of a sphere in plasmonic field", *Phys. Rev. A* **89**, 033841 (2014).
8. A.Y. Bekshaev, K.Y. Bliokh, and F. Nori, "Transverse spin and momentum in two-wave interference", *Phys. Rev. X* **5**, 011039 (2015).
9. M. Neugebauer, T. Bauer, A. Aiello, and P. Banzer, "Measuring the transverse spin density of light", *Phys. Rev. Lett.* **114**, 063901 (2015).
10. M. Neugebauer, J. S. Eismann, T. Bauer, and P. Banzer, "Magnetic and electric transverse spin density of spatially confined light", *Phys. Rev. X* **8**, 021042 (2018).
11. A. Aiello, P. Banzer, M. Neugebauer, and G. Leuchs, "From transverse angular momentum to photonic wheels," *Nature Photon.* **9**, 789 (2015).
12. F.J. Rodríguez-Fortuño, G. Marino, P. Ginzburg, D. O'Connor, A. Martinez, G.A. Wurtz, and A.V. Zayats, "Near-field interference for the unidirectional excitation of electromagnetic guided modes", *Science* **340**, 328 (2013).
13. J. Petersen, J. Volz, and A. Rauschenbeutel, "Chiral nanophotonic waveguide interface based on spin-orbit interaction of light", *Science* **346**, 67 (2014).
14. B. le Feber, N. Rotenberg, L. Kuipers, "Nanophotonic control of circular dipole emission: towards a scalable solid-state to flying-qubits interface", *Nature Commun.* **6**, 6695 (2015).
15. I. Söllner, S. Mahmoodian, S.L. Hansen, L. Midolo, A. Javadi, G. Kirsanske, T. Pregnolato, H. El-Ella, E.H. Lee, J.D. Song, S. Stobbe, and P. Lodahl, "Deterministic photon-emitter coupling in chiral photonic circuits", *Nature Nanotechnol.* **10**, 775 (2015).
16. K.Y. Bliokh, D. Smirnova, and F. Nori, "Quantum spin Hall effect of light", *Science* **348**, 1448 (2015).
17. Y. Lefier and T. Grosjean, "Unidirectional sub-diffraction waveguiding based on optical spin-orbit coupling in subwavelength plasmonics waveguides", *Opt. Lett.* **40**, 2890 (2015).
18. P. Lodahl, S. Mahmoodian, S. Stobbe, A. Rauschenbeutel, P. Schneeweiss, J. Volz, H. Pichler, and P. Zoller, "Chiral quantum optics", *Nature* **541**, 473 (2017).
19. Y. Long, J. Ren, and H. Chen, "Intrinsic spin of elastic waves", *PNAS* **115**, 9951 (2108).
20. C. Shi, R. Zhao, Y. Long, S. Yang, Y. Wang, H. Chen, J. Ren, and X. Zhang, "Observation of acoustic spin", arXiv:1808.03686 (2018).
21. M. Ambati, N. Fang, C. Sun, and X. Zhang, "Surface resonant states and superlensing in acoustic metamaterials", *Phys. Rev. B* **75**, 195447 (2007).
22. C.M. Park, J.J. Park, S.H. Lee, Y.M. Seo, C.K. Kim, and S.H. Lee, "Amplification of acoustic evanescent waves using metamaterial slabs", *Phys. Rev. Lett.* **107**, 194301 (2011).
23. M.-H. Lu, L. Feng, and Y.-F. Chen, "Phononic crystals and acoustic metamaterials", Materials Today **12** (12), 34 (2009).
24. G. Ma and P. Sheng, "Acoustic metamaterials: From local resonances to broad horizons", *Science Advances* **2**, e1501595 (2016).

25. S.A. Cummer, J. Christensen, and A. Alù, "Controlling sound with acoustic metamaterials", *Nature Reviews Materials* **1**, 16001 (2016).
26. I.V. Shadrivov, A.A. Sukhorukov, Y.S. Kivshar, A.A. Zharov, A.D. Boardman, and P. Egan, "Nonlinear surface waves in left-handed materials", *Phys. Rev. E* **69**, 016617 (2004).
27. A.V. Kats, S. Savel'ev, V.A. Yampol'skii, and F. Nori, "Left-handed interfaces for electromagnetic surface waves", *Phys. Rev. Lett.* **98**, 073901 (2007).
28. A.V. Zayats, I.I. Smolyaninov, and A.A. Maradudin, "Nano-optics of surface plasmon polaritons", *Phys. Rep.* **408**, 131 (2005).
29. J. D. Jackson, *Classical Electrodynamics* (Wiley, New York, 1999);
    L.D. Landau, E.M. Lifshitz, and L.P. Pitaevskii, *Electrodynamics of Continuous Media* (Pergamon, New York, 1984).
30. K.Y. Bliokh, A.Y. Bekshaev, and F. Nori, "Optical momentum, spin, and angular momentum in dispersive media," *Phys. Rev. Lett.* **119**, 073901 (2017).
31. K.Y. Bliokh, A.Y. Bekshaev, and F. Nori, "Optical momentum and angular momentum in complex media: from the Abraham-Minkowski de- bate to unusual properties of surface plasmon-polaritons," *New J. Phys.* **19**, 123014 (2017).
32. K. Volke-Sepúlveda, A.O. Santillán, and R.R. Boullosa, "Transfer of angular momentum to matter from acoustical vortices in free space", *Phys. Rev. Lett.* **100**, 024302 (2008).
33. J. Lu, C. Qiu, M. Ke, and Z. Liu, "Valley vortex states in sonic crystals", *Phys. Rev. Lett.* **116**, 093901 (2016).
34. S. Wang, G. Ma, and C.T. Chan, "Topological transport of sound mediated by spin-redirection geometric phase", *Sci. Adv.* **4**, eaaq1475 (2018).
35. A.Y. Bekshaev and K.Y. Bliokh, "Spin and momentum of light fields in dispersive inhomogeneous media with application to the surface plasmon-polariton wave", *Ukr. J. Phys. Opt*. **19**, 33 (2018).
36. A. Aiello and P. Banzer, "The ubiquitous photonic wheel", *J. Opt.* **18**, 085605 (2016).